\documentstyle[11pt,newpasp,twoside]{article}
\def\edcomment#1{\iffalse\marginpar{\raggedright\sl#1\/}\else\relax\fi}
\reversemarginpar

\newcommand{\be}{\begin{equation}}
\newcommand{\ee}{\end{equation}}
 
\begin{document}
\title{DIRECT Distances to Local Group Galaxies}
 \author{Dimitar D. Sasselov}
\affil{Harvard-Smithsonian Center for Astrophysics,\\
60 Garden St., Cambridge, MA 02138, USA}

\begin{abstract}
The two nearby galaxies, M31 and M33, are stepping stones for most of
our current efforts to understand the evolving universe at large
scales. We are undertaking a long term project, called DIRECT,
to improve the direct distance estimate to M31 and M33.
The massive photometry we have obtained as part of our project
over the past 3 years provides us with very good light curves for
known and new Cepheid variables, a large number of eclipsing binaries
and other variable stars.

\end{abstract}

\section{Distances to Local Group Galaxies}

The distances to nearby Local Group galaxies are not known very well,
especially with the new demands of extragalactic stellar astrophysics
which are only likely to increase with the advent 
of several 8-m class telescopes (e.g., Aparicio, Herrero, \&
Sanchez 1998). Nearby galaxies are also crucial calibrators to techniques
for establishing the extragalactic distance scale and determining the value 
of H$_0$ (Mould et al. 1999). Two important spiral galaxies in our nearest
neighbourhood are M31 and M33. Yet, their distances are 
now known to no better than 10-15\%, as there
are discrepancies of $0.2-0.3\;{\rm mag}$ between various distance
indicators (e.g.~Huterer, Sasselov \& Schechter 1995; Holland 1998;
Stanek \& Garnavich 1998) (see Fig.~1).

Direct distances to M31 and M33 are now achieveable, by the use of geometric
techniques, with detached eclipsing binaries and Cepheids, and the use of
8$-$10-m class telescopes for the required spectroscopy. Such direct
distances are the ultimate goal of project DIRECT, which we started
three years ago (Kaluzny et al.~1998, 1999; Stanek et al.~1998, 1999).
The identification of variables suitable for direct study was accomplished
by massive photometry; much of it still continues. In this review I describe
some results from the massive photometry by project DIRECT in M31 and M33.

\begin{figure}[p]

\plotfiddle{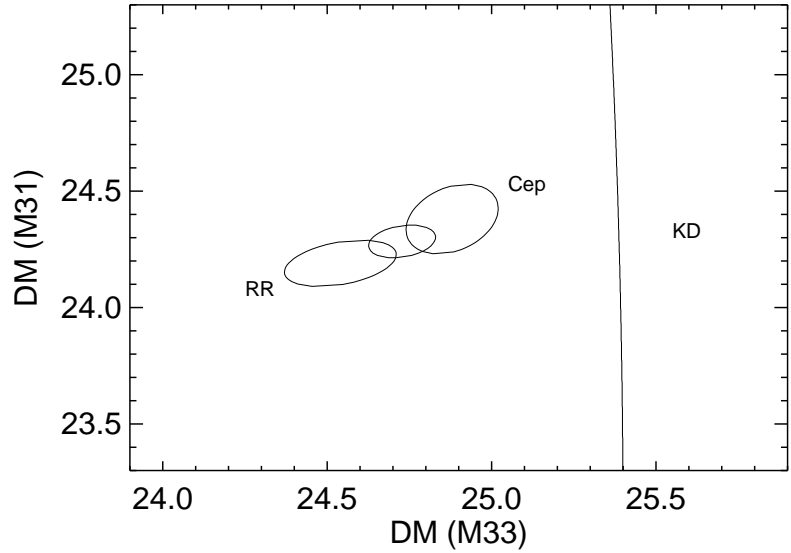}{6.6cm}{0}{65}{65}{-210}{-245}
\caption{Projections of the distance uncertainty ellipsoids for the
nearby galaxies M31 and M33. The RR Lyrae and Cepheid distances are
systematically discrepant in both. The unmarked ellipse is the
projection of the average distance ellipsoid, based on the
simultaneous solution for 15 objects (see Huterer et al.~1995 for
details).  KD stands for a kinematic distance (limit) from VLBI of maser
sources in M33.}

\plotfiddle{sasselovd2.eps}{9.1cm}{0}{45}{45}{-125}{-60}
\caption{The M33 galaxy with superimposed fields observed by the
HST/WFPC2 (small footprints). Also shown are the DIRECT project
$11'\times11'$ fields ABC monitored for variability in 1996 and 1997
(small rectangles) and $23'\times23'$ fields XY monitored in 1998 and
1999.}

\end{figure}

\section{Project DIRECT}

The DIRECT project team now consists of K.~Z.~Stanek (CfA), J. Kaluzny (Warsaw),
A. H. Szentgyorgyi (CfA), J. L. Tonry (Hawaii), L. M. Macri (CfA),
B. J. Mochejska (Warsaw), and myself. Between September
1996 and November 1999 we have obtained $\sim 170$ nights on the
1.2-meter FLWO telescope and 35 nights on the 1.3-meter MDM telescope
for the project, 23 nights on the 2.1-meter KPNO telescope, and four
nights on the Keck II 10-meter telescope. 
We have completely reduced and analyzed data for five of the fields in
M31. We have found in these fields 410 variable stars: 48 eclipsing
binaries, 206 Cepheids and 156 other variables; about 350 of these
variables are newly discovered.  We should stress here that for the
first time detached eclipsing binaries were found in M31 by a CCD
search. We are completing the reduction of the remaining fields and
will continue submitting the next parts of M31/M33 variables catalogs
for publication. In Figure~2 we show the fields observed in M33 by our
project in 1996-97 (small rectangles) and 1998-99 (large rectangles).

The moment our papers are submitted for
publication, all the variable stars lightcurves and finding charts are
made available through the {\tt anonymous ftp} on {\tt
cfa-ftp.harvard.edu}, in {\tt pub/kstanek/DIRECT} directory; and
also through the {\tt WWW} at the {\tt
http://cfa-www.harvard.edu/\~\/kstanek/DIRECT/}.

\section{Results on Cepheids}

So far we have found 206 Cepheids in M31 and 270 in M33. For two we have
obtained Keck II LRIS spectra suitable for detailed abundance analysis and
precise velocities. Our approach has been to obtain very well sampled light 
curves. This secures the identification of all classes of variable
stars; allows meaningful Fourier decompositions for Cepheids  to $i=3$, at
least; and lets us deal with some issues of metallicity differences and 
blending in the Cepheid population.

For example, pulsation resonances are known to play an important role in 
shaping the morphology of the light curves of Cepheids, e.g. the
Hertzsprung bump progression centered on period of 10$^d$,
identified with the 2:1 resonance 2$\omega_0 \approx \omega_2$.
Not surprisingly, it is very sensitive to the
opacity in the Cepheid envelope, a crude sort of asteroseismology, and is
a good indicator of the metallicity of a Cepheid sample (Buchler 1997).
Thanks to data from the microlensing surveys the metallicity dependence of
the resonance center has now been established from Cepheids in the Galaxy,
LMC, and SMC (Beaulieu \& Sasselov 1997). The resonance center is shifted
by full 2 days for a $\Delta$[Fe/H]=0.7. For such purposes
the Fourier technique is entirely model and photometry independent, it
only depends on well sampled light curves in an optical band.

The two populations of Cepheids in our M31 data, which we expect to be of
different metallicity, are located in two spiral arms. Although
our $inner$ arm sample is still small (20), its resonance center is
defined and shifted by $\sim$1 day in the fashion described above.
The resonance troughs are filled with outliers in Fields B\&C,
due to the sample of high-metallicity Cepheids from the inner arm.
The galactocentric distances of the two samples are 4 and 10~$kpc$,
respectively. At these distances the abundances in the M31 disk are close
to solar and slightly higher.

\begin{figure}[p]

\plotfiddle{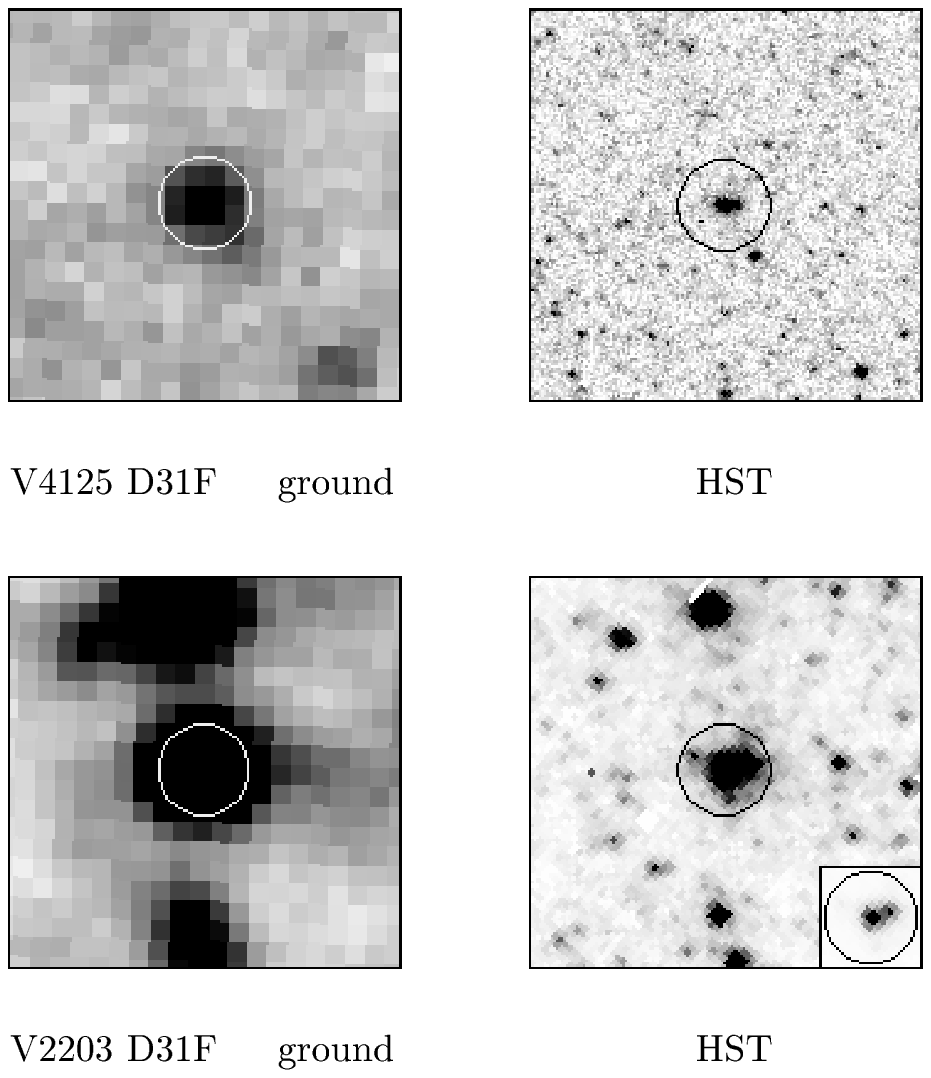}{9.5cm}{0}{90}{90}{-165}{-345}
\caption{Two examples of blending of Cepheids in M31 found by our
current work on DIRECT and archival {\em HST} data.}

\plotfiddle{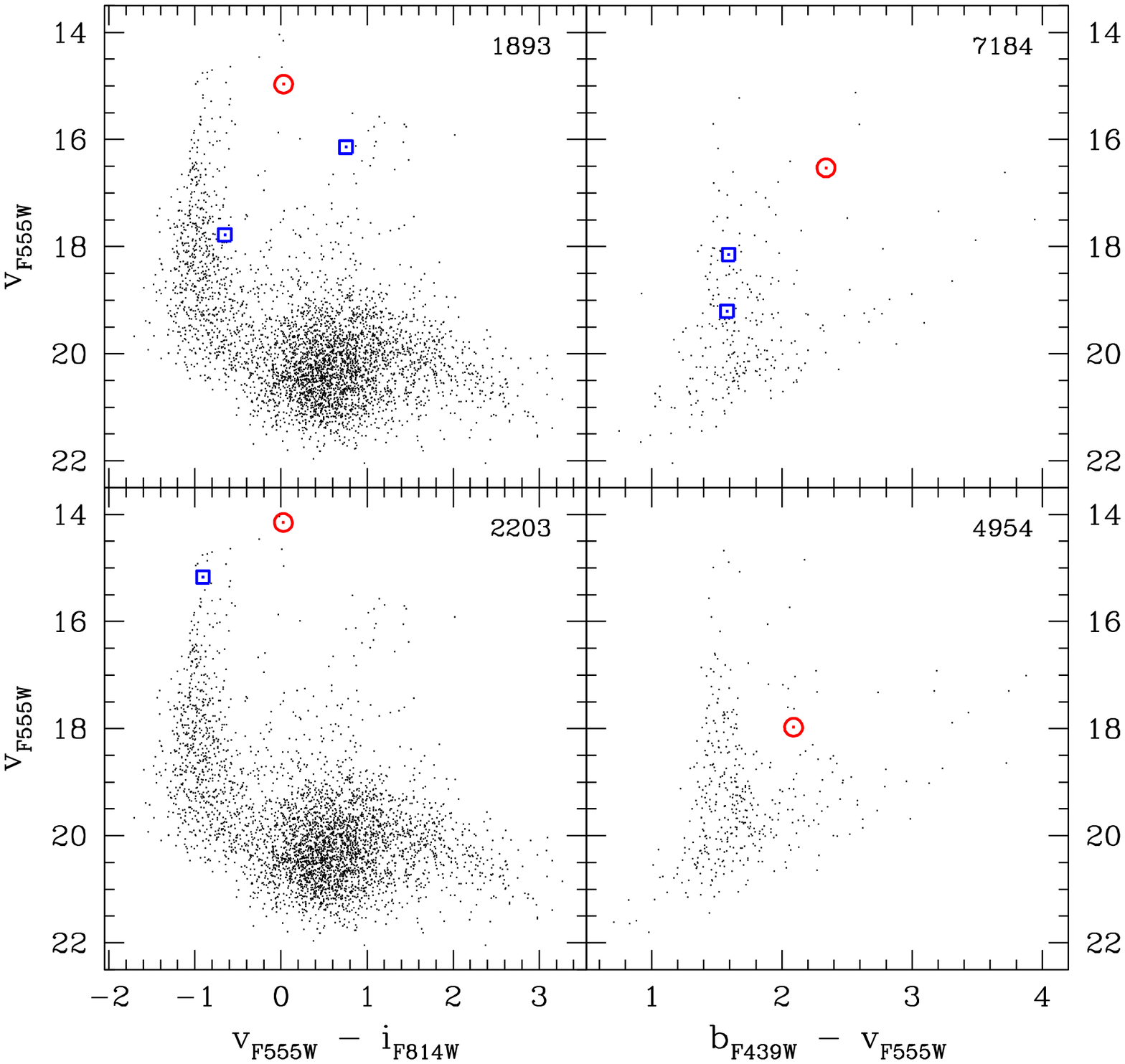}{8cm}{0}{38}{38}{-105}{-45}
\caption{The color-magnitude diagrams for 4 of the Cepheid blends in
M31. In each the Cepheid is marked with a circle.}
\end{figure}

\section{Correction for Blending}

We are using archival $HST$ images coinciding with our fields (see Fig.~2)
to study the environments of the Cepheids discovered by the ground-based
observations of project DIRECT. We find that roughly 50\% of our Cepheids
are blended with other stars which contribute more than 10\% of the light
(in V-band). Cepheid blending can account for up to 40\% of the measured flux
from some of the Cepheids as found in Mochejska et al. (1999), where we
report on the sample of 20 Cepheids in M31. The analysis of 90 Cepheids in
M33 is practically complete. We find that the average (median)
V-band flux contribution from luminous companions which are not
resolved on the ground-based images is about 19\% (12\%) of the flux
of the Cepheid in M31 and 27\% (17\%) in M33. The average (median)
I-band flux contribution is about 34\% (24\%) in M33, i.e. there appears
to be no bias in the color of the blends.

Our indirect distances to the two galaxies will be easily corrected for this
blending. Note that Cepheid V7184 in Fig.~4 was discovered
to be a blend by analysing our ground-based DIRECT light curve, before
we had $HST$ data on it (Kaluzny et al. 1998). Cepheid V4954 was flagged 
accordingly. This shows the usefulness of good light curves, albeit limited 
to strong blending only.

      Our ground-based resolution in M31 and M33 corresponds to the
HST resolution at about 10 Mpc. Blending leads to systematically low distances 
to galaxies observed with the HST, and therefore to systematically high
estimates of H$_0$. We predict the Cepheid blending effects for a galaxy
at $\sim$25 Mpc observed by HST to be severe with the corresponding implications
to the extragalactic distance scale.

{\bf Acknowledgements.} This work was supported in part by NSF grant
AST-9970812.

\end{document}